\documentclass[10pt, final, journal, a4paper, twocolumn]{IEEEtran}
\usepackage[dvips]{graphicx}
\usepackage{times}
\usepackage{color}
\usepackage{mathrsfs}
\usepackage{amsmath}
\usepackage{amssymb}
\usepackage{cite}
\usepackage{subfigure}
\begin{document}
%
\title{Utility-Based Wireless Resource Allocation for Variable Rate Transmission}

\author{Xiaolu~Zhang,~\IEEEmembership{Student Member,~IEEE,}
        Meixia~Tao,~\IEEEmembership{Member,~IEEE,}
        and~Chun~Sum~Ng,~\IEEEmembership{Member,~IEEE}
\thanks{Manuscript received March 21, 2007; revised July 19, 2007;
        accepted September 17, 2007. The editor coordinating the
        review of this paper and approving it for publication is
        D. Wu. The paper was presented in part
        at the IEEE International Conference on Communications, Glasgow,
        June 2007.
        }
\thanks{The authors are with the Department of Electrical and Computer Engineering,
National University of Singapore, Singapore 117576, (e-mail:
zhangxiaolu@nus.edu.sg; mxtao@nus.edu.sg; elengcs@nus.edu.sg).}}
 \maketitle
\begin{abstract}
For most wireless services with variable rate transmission, both
average rate and rate oscillation are important performance
metrics. The traditional performance criterion, utility of average
transmission rate, boosts the average rate but also results in
high rate oscillations. We introduce a utility function of
instantaneous transmission rates. It is capable of facilitating
the resource allocation with flexible combinations of average rate
and rate oscillation. Based on the new utility, we consider the
time and power allocation in a time-shared wireless network. Two
adaptation policies are developed, namely, time sharing (TS) and
joint time sharing and power control (JTPC). An extension to
quantized time sharing with limited channel feedback (QTSL) for
practical systems is also discussed. Simulation results show that
by controlling the concavity of the utility function, a tradeoff
between the average rate and rate oscillation can be easily made.
\end{abstract}
\IEEEpeerreviewmaketitle

\section{Introduction}
An important aspect of wireless systems is dynamic channel
characteristics. One promising approach for addressing this issue
is to dynamically allocate limited resources based on channel
information and system preferences. Traditional investigations on
wireless resource allocation pay much attention to hard real-time
services. Therein, the goal is to smooth out channel variation and
build ``bit pipes'' that deliver data at a fixed rate. The rapid
growth of the Internet has led to an increasing demand for
supporting transmissions of best-effort service in wireless
systems. These applications allow variable-rate transmission and
are tolerant of high rate oscillations. Therefore, opportunistic
communications~\cite{ViswanathTL02j} have been introduced to
achieve higher system throughput. The concept of opportunistic
communications is essentially to transmit more information in good
channel states and less in poor ones. Hard real-time service and
best-effort service may be viewed as two extremes of
rate-oscillation sensitivity. However, services such as many audio
and video applications generally expect a balance between average
rate and rate oscillation. If constant-rate transmission
algorithms are used, the transmission efficiency would be very
low. On the other hand, opportunistic scheduling schemes, such as
\cite{JalaliPP00c} and\cite{Stolyar05j}, whose objective is to
maximize a utility of average rates, can improve efficiency in
terms of average rate but result in high oscillation in
instantaneous transmission rates. This thus motivated the need for
a new criterion that can be used to facilitate the choice of the
combinations of average rate and rate oscillation.

In this letter we propose a new network objective function,
namely, Time-average Aggregate concave Utility of instantaneous
transmission Rate (TAUR). To illustrate the underlying mechanism
of the proposed objective function, let us consider transmitting a
same data stream using two different schemes. For scheme one, the
data stream is transmitted at a constant speed of $1$ Mbit/s
during the interval of $10$ seconds. For scheme two, no data is
transmitted in the first 9 seconds and $10$ Mbit/s is used for
transmission in the last second. Obviously, the utilities of the
two transmission schemes are identical if the utility is defined
as a function of average transmission rate. However, the
time-average concave utility as a function of instantaneous
transmission rate for scheme one is higher than that for scheme
two, which is expected if the degree of user satisfaction is
concerned. Thus, the resource allocation based on TAUR should be
able to balance the average rate and rate oscillation over time by
adjusting the concavity of the utility function.

The TAUR-based resource allocation problems are studied in a
multi-user wireless system in an adaptive time-division fashion.
We first consider the optimal time sharing (TS)-based scheduling
policy in a backlogged system with constant power allocation.  For
a strictly concave utility, our analysis shows that the TS policy
allows users with relatively better channel conditions to share a
same time frame. We then propose a joint optimal time sharing and
power control (JTPC) strategy where both the time-sharing fraction
and the transmit power can be varied over time. In addition, a
quantized TS policy with limited channel feedback (QTSL) is
proposed for the ease of practical implementation.

\renewcommand{\vec}[1]{\mbox{\boldmath$#1$}}
\section{System Model}
We consider a single cell consisting of $N$ mobile users
communicating with a common base station. The communication link
between each user and the base station is modelled as a slowly
time-varying fading channel with additive white Gaussian noise
(AWGN). The channel coefficients remain approximately unchanged
during each time frame, but can vary from one frame to another.
Let the instantaneous channel gain of user $i$ at any given time
frame $t$ be denoted by $g_{i}(t)$. The network channel gain is
denoted by the $N$-tuple $\mathbf{g}(t) \triangleq
(g_{1}(t),g_{2}(t),\ldots,g_{N}(t))$, and has a joint probability
density function (PDF) $f(\mathbf{g})$. Let $p_i(t)$ denote the
transmit power allocated to or from user $i$. The achievable
transmission rate of user $i$ in the absence of other users can be
expressed as~\cite{QiuC99j}
\begin{equation}\label{SNR gap}
c_{i}(t)=\log_{2}\left[1+\frac{p_i(t) g_{i}(t)}{\beta
N_{0}}\right],
\end{equation}
where $N_{0}$ is the noise power, and $\beta$ is the
signal-to-noise ratio (SNR) gap~\cite{QiuC99j}. We assume that
each time frame can be accessed by all the $N$ users in an
adaptive time-sharing fashion. Let $\vec{\rho}({\mathbf
g})=(\rho_1, \rho_2, \ldots, \rho_N)$ denote the time-sharing
adaptation policy with respect to the network channel gain
$\mathbf{g}$, where $\rho_i$ represents the fraction of the frame
duration allocated to user $i$. Without loss of generality, the
interval of a time frame is normalized. The actual transmission
rate of user $i$ at the $t$-th time frame, $r_i(t)$, is given by
$r_i(t)=\rho_{i}c_{i}(t)$. The frame index $t$ in $r_i(t)$ and
$g_i(t)$ may be omitted hereafter if no confusion occurs.

The utility considered here is a function of the instantaneous
transmission rate. For user $i$, we denote its utility as
$U_i(r_i(t))$. The exact expression for the utility $U_i(\cdot)$
is not crucial. The analysis throughout this paper is valid for
any utility function that is increasing, differentiable and
concave.

\section{Time Sharing}\label{section TA}
In this section, we assume that the transmission powers between BS
and mobiles are constant and identical for different users, i.e.,
$p_i(t)=p,\forall i, t$, and that the wireless network is fully
loaded. We choose the aggregate utility, which is the sum of
individual user utilities, as the performance measure. The goal is
to find the optimal time-sharing adaptation policy
$\vec{\rho}^*({\mathbf g})$ relative to the instantaneous network
channel condition $\mathbf g$, so as to maximize the TAUR of the
system. Since the channel processes are ergodic, the optimization
problem can be expressed mathematically as
\begin{eqnarray}\label{TS_o}
\max_{\vec{\rho}({\mathbf g})}~~I_{\rm TS}& \triangleq
&\mathbb{E}_t\left[\sum^N_{i=1} U_{i}(r_{i}(t))\right]\\
&=&\int_{\mathbf{g}} \sum^N_{i=1} U_{i}(\rho_{i}(\mathbf{g}), g_i)
f(\mathbf{g}) \mbox{d}\mathbf{g} \nonumber\\
\mbox{s.t.} &&\sum^N_{i=1}\rho_{i}(\mathbf{g})=1
\label{constraint2}.
\end{eqnarray}
where notation $\mathbb{E}_t[\cdot]$ represents the time average.

Since the constraint~(\ref{constraint2}) is defined for all
channel states, the average aggregate utility maximization in
(\ref{TS_o}) is equivalent to maximizing the instantaneous
aggregate utility for every channel state. Furthermore, since the
utility $U_i(\cdot)$ is a concave function of $r_i$ by assumption,
$U_i(\cdot)$ is also concave in $\rho_i$. Therefore, taking the
derivative of the Lagrangian $\sum^N_{i=1}
U_{i}(\rho_{i}(\mathbf{g}), g_i)+\lambda
[1-\sum_{i=1}^N\rho_i(\mathbf{g})]$, and equating it to zero, we
obtain $\rho_i^*$ as
\begin{equation}\label{optimal rho solution}
\rho_{i}^*(\mathbf{g})=\left[\left(\frac{\partial U_{i}(\rho_i,
g_i)}{\partial\rho_{i}}\right)^{-1}\left(\lambda\right)\right]^+,
~~i=1,2,\ldots,N.
\end{equation}
In~(\ref{optimal rho solution}), $\left({\partial
U_{i}}/{\partial\rho_{i}}\right)^{-1}(\cdot)$ is the inverse
function of $({\partial
U_{i}}/{\partial\rho_{i}})(\cdot)$\footnote{When the utility
function is strictly concave, $({\partial
U_{i}}/{\partial\rho_{i}})(\cdot)$ is a monotonically decreasing
function of $\rho$ and, hence, its inverse exists.}, and $(x)^+
\triangleq \max (0,x)$. The Lagrange multiplier $\lambda$ can be
determined using the constraint (\ref{constraint2}).
 As can be seen, the
optimal time-sharing policy is only a function of the
instantaneous channel conditions and is independent of the channel
statistics. The explicit solution of the proposed optimal TS
policy in a two-user network with log utility is discussed in
\cite{ZhangTN07icc}.

We now compare the proposed TS policy for maximizing the
time-average aggregate utility of instantaneous rate with the
existing gradient scheduling (GS) policy~\cite{Stolyar05j} for
maximizing the aggregate utility of average transmission rate. As
shown in~\cite{Stolyar05j}, the GS policy maximizes the weighted
sum of instantaneous rates in the system. In a time-shared
wireless network, this results in choosing the user satisfying the
following condition to transmit during the whole time frame:
\begin{equation}\label{eqn:GS}
i^*(t)=\arg \max_{1\le i\le N} {\frac{\partial
U_i(R_i(t))}{\partial R_i(t)} c_i(t)}.
\end{equation}
Here, $R_i(t)$ is updated as $R_i(t)=(1-\alpha)R_i(t-1)+\alpha
c_i(t-1)$ for $i=i^*(t-1)$ and $R_i(t)=(1-\alpha)R_i(t-1)$ for
$\forall i\ne i^*(t-1)$, with arbitrary initial value $R_i(0)$,
and $\alpha>0$ is a fixed small parameter.
On the other hand, in the proposed TS policy, it shall be clear in
Section V that the decreasing marginal utility gives opportunity to
the users in poor channel condition to share the time frame.

Fig.~\ref{fig compare with MAXRATE} shows the simulated TAUR in
the network by using the proposed TS policy and the existing GS
policy. Here the utility of both policies is specified to have the
form
\begin{equation}\label{equ_utility_function}
U(r)=\ln\left(1+\frac{r}{A}\right),
\end{equation}
where $A >0$ is a concavity indicator. The channels are assumed to
be Rayleigh fading and the SNR gap is set to $8.2$ dB for all
users which corresponds to a bit-error-rate requirement of
$10^{-5}$ when adaptive quadrature amplitude modulation (QAM) is
used. The number of users in the network varies from 8 to 32 and
their channel conditions are symmetric. The concavity indicator
$A$ is set at $0.1$. Fig.~\ref{fig_mean_throughput} compares the
mean and the standard deviation of the transmission rate achieved
by TS and GS policies for $A=0.1, 1$ and 10 with $N=32$. The
performances of the GS policy at different values of $A$ are
identical due to the channel symmetry among the users. In fact, it
is an extreme case of our TS policy when $A \to \infty$ . Although
GS can obtain the maximum average rate, the standard deviation of
the rate increases rapidly as the average SNR increases. The TS
policy, on the other hand, has a flexible balance between the
average rate and the rate oscillation through adjusting the
concavity of the utility function.

\section{joint time-sharing and power control}\label{section JTPA} %
In this section we allow both the transmission time and power to
change with respect to channel conditions in each time frame. The
optimization problem in Section~\ref{section TA} is extended to
finding the joint optimal time-sharing and power control policy.
Uplink and downlink transmission are considered separately due to
different power constraints. For the uplink, the power source is,
generally, rechargeable batteries attached to the mobile devices.
Thus, the optimization is subject to each user's average power
constraint. Mathematically, this can be represented as
\begin{eqnarray}\label{equ_JTPC_objective}
\max_{(\vec{\rho},\mathbf{p})}& & I_{\rm JTPC} \triangleq
\int_{\mathbf{g}}\sum^N_{i=1}
U_{i}(\rho_{i}(\mathbf{g}),p_{i}(\mathbf{g}),g_{i})f(\mathbf{g})\mbox{d}\mathbf{g}\\
\label{eqn_JTPC_2}
 {\rm s.t.} & & \sum^N_{i=1}\rho_{i}(\mathbf{g})=1\\
\label{equ_constraint3}
&&\int_{\mathbf{g}}\rho_{i}(\mathbf{g})p_{i}(\mathbf{g})f(\mathbf{g})\mbox{d}\mathbf{g}=\bar{p}_{i},~i=1,2,\ldots,N,
\end{eqnarray}
where $\bar{p}_{i}$ is the average power constraint of user $i$.

Note that the utility function $U(\rho, p, g)$ is concave in
$\rho$ and $p$ separately based on our assumption, but not in both
$\rho$ and $p$. Moreover, the equality constraints in
(\ref{equ_constraint3}) are nonlinear. To make the problem more
tractable, we define $s=p\rho$. It can be shown that $U(\rho, s,
g)$ is concave in both $\rho$ and $s$ (since its Hessian matrix is
negative semidefinite). The problem thus falls into the
 classic calculus of variations~\cite{GottfriedW73b}.
Applying the Euler-Lagrange equation results in the
following necessary and sufficient conditions for the optimal
solution $\rho_i^*$ and $s_i^*$:
\begin{eqnarray}
\label{equ_JTPC_constraint1}&&\frac{\partial U_{i}}{\partial s_{i}}+\lambda_{i}=0,~~~i=1,2,\ldots,N\\
\label{equ_JTPC_constraint2}&&\frac{\partial U_{i}}{\partial
\rho_{i}}f(\mathbf{g})+\lambda_{0}(\mathbf{g})=0,~~~i=1,2,\ldots,N
\end{eqnarray}
where $\lambda_{0}(\mathbf{g})$ and $\lambda_i$ are Lagrange
multipliers and determined by constraints (\ref{eqn_JTPC_2}) and
(\ref{equ_constraint3}). The closed-form solutions to the above
equations are generally difficult to obtain due to nonlinearity of
the utility function $U$ in $\rho$ and $s$. The nonlinear
Gauss-Seidel algorithm~\cite{BertsekasT97b} can be used to search
for the optimal time-sharing vector $\vec{\rho}^*$ and vector
$\mathbf{s}^*=[s_1^*,s_2^*,\ldots,s_N^*]$ under the average power
constraint $\mathbf{\bar{p}}=[\bar{p}_1,\bar{p}_2, \ldots,
\bar{p}_N]$. It is outlined as follows.

{\small
\begin{enumerate}
    \item{\emph{Initialize $\mathbf{s}$}}: set $ \mathbf{p}^{(0)} = N\mathbf{\bar{p}}$,
        $\rho_i=1/N$, $\forall i$, let $j=0$ and calculate the
        initial $\mathbf{s^{(0)}}$.
    \item{\emph{Update $\mathbf{\rho}^{(j)}$ }}: given $\mathbf{s}^{(j)}$, find the optimal time-sharing vector
    $\vec{\rho}^{(j)}$ using (\ref{equ_JTPC_constraint2})
    \[
        \rho_{i}^{(j)}(\mathbf{g})=\left[\left(\frac{\partial
        U_{i}}{\partial\rho_{i}}\right)^{-1}\left(\lambda_0',s^{(j)}_i\right)\right]^+,
        ~~i=1,2,\ldots,N ,
    \]
    where $\lambda_0'$ satisfies (\ref{eqn_JTPC_2})
    \item{\emph{Compute the average aggregate utility $I^{(j)}$ using (\ref{equ_JTPC_objective})}}
    \item{\emph{Update $\mathbf{s}^{(j+1)}$}}: given $\vec{\rho}^{(j)}$, find the optimal power control vector
    $\mathbf{s}^{(j+1)}$ using (\ref{equ_JTPC_constraint1})
        \[
            s_{i}^{(j+1)}(\mathbf{g})=\left[\left(\frac{\partial
            U_{i}}{\partial
            s_{i}}\right)^{-1}\left(\lambda_i,\rho^{(j)}_i\right)\right]^+
            ~\mbox{for}~i=1,2,\ldots,N,
        \]
    where $\lambda_i$ satisfies (\ref{equ_constraint3}) for $i=1,2,\ldots,N$ and let $j=j+1$.
 \item{\emph{Repeat Steps 2)-4) until $I^{(j+1)}-I^{(j)}<\Delta$, where $\Delta$ is a small number. }}
\end{enumerate}}
At each iteration, the optimization of $\vec{\rho}$ and $\mathbf
s$ is carried out successively. Steps 2) and 4) involve only the
calculation of a one-dimensional maximization problem whose
solution is given in Section~\ref{section TA}. The condition that
$U$ is continuously differentiable and concave in ($\rho$, $s$)
guarantees the convergence of the nonlinear Gauss-Seidel
algorithm. The proof can be seen in~\cite[Prop 3.9 in Section
3.3]{BertsekasT97b}.

The problem formulation for the downlink differs from the uplink
only in the power constraint, which is given by $
\int_{\mathbf{g}}\sum^N_{i=1}\rho_{i}(\mathbf{g})p_i(\mathbf{g})f(\mathbf{g})\mbox{d}\mathbf{g}=\bar{p}
$. A similar problem-solving approach to the one proposed for the
uplink can also be obtained and hence is omitted.

Although JTPC utilizes two degrees of freedom in resource
allocation and has much higher computational complexity, its
performance is not expected much higher than that of TS in the
high SNR region. This is attributed to the fact that the
transmission rate is linear in time, but concave in transmission
power. That is, at high SNR, the gain from power control is
smaller than from time sharing adaptation. This is also verified
by the simulation results shown in Fig.~\ref{fig_TS_JTPC}, where
we compare the average aggregate utilities obtained by the
proposed TS and JTPC schemes. It is observed that at high SNR the
performance gain of JTPC over TS is not noticeable. Hence, in
Section \ref{sec_implementation}, we assume the absence of power
control.

\section{Implementation Issues}\label{sec_implementation}
We have provided the analytical results for TS and JTPC policies. In
this section, we address some important implementation issues,
including quantized time sharing fractions, limited channel
feedback, and fairness.

\subsection{Quantized Time Sharing
With Limited Channel Feedback}\label{section_QTSL} In scheduling
the downlink transmission, the BS needs to know each user's
channel state information (CSI). This could be gained by sending
the CSI from each user to the base station through a feedback
channel upon channel estimation at each user terminal. In
practice, perfect channel feedback is not feasible due to limited
capacity of the feedback links. We assume in this subsection that
the channel estimate of each user is quantized into $K=2^M$
regions using $M$ bits. Let $\mathbb{G}=\{G_1,G_2,\ldots,
G_{K+1}\}$ be the set of channel gain thresholds in increasing
order with $G_1=0$ and $G_{K+1}=\infty$. If the channel gain of
user $i$ falls into range $[G_k,G_{k+1})$, we say user $i$ is in
channel state $k$, and denote it as $S_i=k$. Suppose we apply the
equal-probability method to do the channel partitioning and the
channel gains follow exponential distribution, the threshold set
$\mathbb{G}$ can be determined easily.

Furthermore, time sharing fractions in practice cannot be an
arbitrary number, but are restricted to a finite set of values due
to switching latency and difficulties in rigid synchronization.
Therefore, we assume that a time frame is partitioned into $L$
slots with equal length. Correspondingly, the number of users
which can transmit in the same frame is limited by $L$. At the
beginning of each frame, the base station computes the optimal
time-sharing vector $\vec{\rho}^*$ defined
in~(\ref{eqn_optimization_problem_limit_feedback}) upon obtaining
the network channel states $\mathbf{S}=[S_1,S_2,\ldots,S_N]$.
\begin{figure*}[!t]
\normalsize \setcounter{equation}{11}
\begin{eqnarray}\label{eqn_optimization_problem_limit_feedback}
\vec{\rho}^*(\mathbf{S})&=&\mbox{arg} \max_{\vec{\rho}(\mathbf{S})}~\int_{G_{S_1}}^{G_{S_1+1}}\int_{G_{S_2}}^{G_{S_2+1}}\ldots \int_{G_{S_N}}^{G_{S_N+1}}\left[\sum_{i=1}^N  U_i(\rho_i(\mathbf{g}),g_i)\right]\mbox{d}\mathbf{g}\\
\mbox{s.t.}&&\rho_i \in
\left\{0,\frac{1}{L},\frac{2}{L},\ldots,1\right\},~\forall
i\in\{1,\ldots, N\}~\mbox{and}~\sum_{i=1}^N \rho_i =1.\nonumber
\end{eqnarray}
\setcounter{equation}{12} \hrulefill
\end{figure*}

The time sharing policy considered here maps the current channel
states $\mathbf{S} \in \mathbb{R}_+^N$ to a time-sharing vector To
avoid the exponential complexity in exhaustive search, an online
greedy algorithm with complexity of $\mathcal{O}(LN)$ is proposed.
Beginning with an initial solution $\vec{\rho}=[0,0,\ldots,0]$,
each time slot is assigned at one iteration to the most favorable
user that maximizes the increment of the current objective till
the total $L$ slots are traversed. The greedy algorithm is
outlined below:

{\small
\begin{enumerate}
\item \emph{Initialization}\\
Let $v=0$ (the index of the time slot), $\rho_i^{(0)}=0$ and $U_i^{(0)}=0~(\forall i\in\{1,\ldots,N\})$.%
 \item \emph{Allocate the $(v+1)^{th}$ time slot to the user indexed by
 ${i}^*$}
\begin{equation}\label{eqn_greedy}
i^*=\mbox{arg}~\max_{i\in\{1,\ldots,N\}}~\int_{G_{S_i}}^{G_{S_i+1}}\bigg[U_i(\rho_i^{(v)}+1/L)-U_i(\rho_i^{(v)})\bigg]
\mbox{d}{g_i}.
\end{equation}
Let $\rho_{i^*}^{(v+1)}=\rho_{i^*}^{(v)}+1/L$ and $\rho_{i}^{(v+1)}=\rho_{i}^{(v)}$ for $i\neq i^*$.%
\item \emph{Let $v=v+1$, and return to Step 2) until $v=L$}
\end{enumerate}
}

 It is shown in the appendix that this algorithm leads to the
optimal solution to
Problem~(\ref{eqn_optimization_problem_limit_feedback}).

Fig.~\ref{fig_QTSL1} shows the performance of the quantized time
sharing with limited channel feedback (QTSL) policy. The CSI is
quantized at 3-bit resolution, and the number of time slots is the
same as the number of users in the network. It is seen that the
performance of QTSL approaches that obtained by the
optimal TS policy. 
%
In Fig.~\ref{fig_QTSL2}, we illustrate the performance of the QTSL
policy when the number of time slots in a time frame is half of the
number of users in the network. This time we also vary the number of
channel feedback bits from $1$ to $3$. There is a performance gap
between QTSL and TS, but the average aggregate utility obtained by
QTSL with $N=32$ and $L=16$ is still higher than that of the optimal
TS policy with $N=16$. In addition, the performance gain is limited
when the CSI is quantized using more than two bits.

\subsection{Fairness Guarantee} \label{section adaptive weight
adjustment} The scheduling schemes based on the aggregate utility
maximization developed in previous sections may not guarantee
fairness for users with different channel statistics, such as path
loss and shadowing. This is because the TS scheme tends to
allocate more time to the user with higher average SNR. A commonly
used fairness criterion in computer networks is the max-min
fairness~\cite{Jaffe81j}. Unlike wireline networks, the wireless
network suffers from time-varying channel impairments. Thus,
time-average utility max-min (TUMM) fairness, defined below, is
meaningful in wireless networks.

\emph{Definition 1}: A time sharing policy $\vec{\rho}$ is
time-average utility max-min fair if for each $i\in\{1,
2,\ldots,N\}$ and any other time sharing policy $\bar{\vec{\rho}}$
for which $\mathbb{E}[U_i(\rho_i)]<\mathbb{E}[U_i(\bar{\rho_i})]$,
there exists some $j$ with
$\mathbb{E}[U_i(\rho_i)]\geq\mathbb{E}[U_j({\rho_j})]>\mathbb{E}[U_j(\bar{\rho}_j)]$,
i.e., \emph{increasing some component $\mathbb{E}(U_i(\rho_i))$
must be at the expense of decreasing some already smaller
component $\mathbb{E}[U_j(\bar{\rho_j})]$}

Since the resource allocation based on TUMM fairness does not
concern the fairness at any instant, it allows the scheduler to
exploit the instantaneous fluctuation of the channel conditions. We
formulate the TUMM-based time sharing as a multiple-objective
programming problem:
\begin{eqnarray}
\max_{\vec{\rho}(\mathbf{g})}&& a\\
\label{eqn_TUMM_constaint}
\mbox{s.t.} && \mathbb{E}[U_i(\rho_i(\mathbf{g}),g_i)]= a,~i=1,2,\ldots,N\\
            &&\sum^N_{i=1}\rho_{i}(\mathbf{g})=1,\nonumber
\end{eqnarray}
 where $a$ is a variable to be maximized. When $U_i(\cdot)$'s
 are all strictly concave, there exists a unique solution.

 A frequently used method in multiple objective
programming problem is the point estimate weighted-sum
approach~\cite{Steuer86b}. In this method, each objective is
multiplied by a weight $w_i$. Then, the $N$ weighted objectives
are summed to form a weighted-sum objective function, denoted as
$\sum^N_{i=1} w_i \mathbb{E}[U_i(\rho_i)]$. The weights $w_i$'s
are chosen such that (\ref{eqn_TUMM_constaint}) is satisfied. It
can be proven that the solution to maximize the average aggregate
weighted utility is also Pareto-optimal. Take the two-user case
for example. The notion of fairness can be realized by dynamically
adjusting the weights, i.e., dynamically adjusting the moving
direction towards the intersection of the Pareto-optimal frontier
and the straight line $\mathbb{E}[U_1]=\mathbb{E}[U_2]$ while
keeping the two users' average utility on the Pareto-optimal
frontier. This weight adaptation method can be used to guarantee
fairness when the users have different channel distributions.

\section{Conclusion}\label{section conclusions}
We develop a new framework for resource allocation in wireless
networks for variable-rate transmission. The time-average
aggregate utility of instantaneous transmission rate is proposed
to jointly optimize the resulting average rate and rate
oscillation. In particular, a time-sharing policy and a joint
time-sharing and power control policy are designed to exploit the
channel fluctuation. The effects of partial channel state
information and discrete time sharing fractions are also studied.
Furthermore, an adaptive method to guarantee strict fairness among
users with different channel statistics is discussed.

\appendix[Optimality Proof of The Greedy
Algorithm] {\small  Consider one realization of $\vec{S}$ and
$\rho_i\in\{0,{1}/{L},{2}/{L},\ldots,1\}$. For simplicity, we
define $ \tilde{U}_i(\rho_i):=\int_{G_{S_i}}^{G_{S_i+1}}
U_i(\rho_i(\mathbf{g}),g_i)\mbox{d}{g_i}$ and $
d_i(\rho)=\tilde{U}_i(\rho)-\tilde{U}_i\left(\rho-{1}/{L}\right)$.
Let $D_L$ denote the set of $L$ largest elements in $
D=\{d_i(\rho)|i=1,2,\ldots,N, \rho= 0, 1/L, 2/L,\ldots 1\}. $
Then, the Lagrangian can be written as:
\begin{equation}\label{eqn_Largranian_function}
\sum_{i=1}^{N}(\tilde{U}_i(\rho_i)-\lambda
\rho_i)=\sum_{i=1}^{N}[\tilde{U}_i(0)+(d_i(1/L)-\lambda)+\ldots+(d_i(\rho_i)-\lambda)].
\end{equation}
Let $\lambda$ be the smallest $d_i(\rho)$ in $D_L$, then
\begin{eqnarray}
 d_i(\rho)-\lambda \left\{ \begin{array}{ll}
 \geq 0& \mbox{if}~d_i(\rho)\in D_L\\
 < 0& \textrm{otherwise}
\end{array}\right.. \end{eqnarray}
Therefore the Lagrangian (\ref{eqn_Largranian_function}) is
maximized by
\begin{eqnarray}\label{eqn_rho_App}
 \rho_i^*=\left\{ \begin{array}{ll}
 0& \mbox{if}~d_i(1/L)\notin D_L\\
1& \mbox{if}~d_i(1)\in D_L\\
\rho& \mbox{if}~d_i(\rho)\in D_L~\mbox{and}~d_i(\rho+1/L)\notin
D_L.
\end{array}\right..\end{eqnarray}
Due to the concavity of the utility function, we have
 $d_i(1/L)>d_i(2/L)>\ldots>d_i(N_i/L)>d_i((N_i+1)/L)>\ldots>d_i(1)$.
Here suppose that $ d_i(N_i/L)\in D_L$ and $ d_i((N_i+1)/L) \notin
D_L$, then $\sum_i N_i = |D_L|$. Since $\rho_i^* = N_i/L$ from
(\ref{eqn_rho_App}), $\sum_i \rho_i^*=|D_L|/L=1$ holds for the
chosen $\lambda$. $\rho_i^*$ is an optimal solution of Problem
(\ref{eqn_optimization_problem_limit_feedback}).

Since Step 2) computes the $L$ largest $d_i(\rho)$ in the
decreasing order of their values, the greedy algorithm obtains the
optimal solution. }

\bibliographystyle{IEEEtran}
\bibliography{bibliography}
\newpage
\begin{figure}
\centering
\includegraphics[width=3.3in]{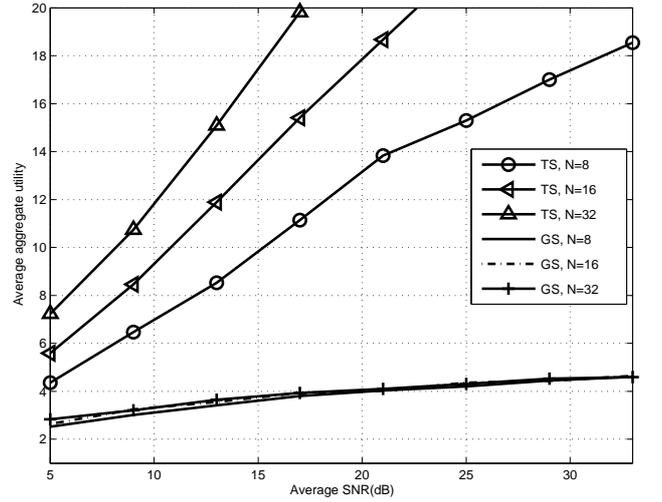}
\caption{Performance comparison of different resource allocation
schemes} \label{fig compare with MAXRATE}
\end{figure}

\begin{figure}
\centering
\includegraphics[width=3.3in]{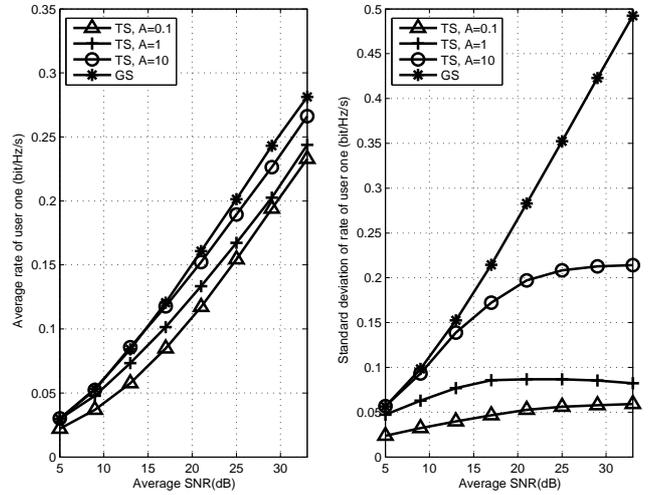}
\caption{ Average rate and standard deviation of rate versus
average SNR using TS and GS} \label{fig_mean_throughput}
\end{figure}

\begin{figure}
\centering
\includegraphics[width=3.3in]{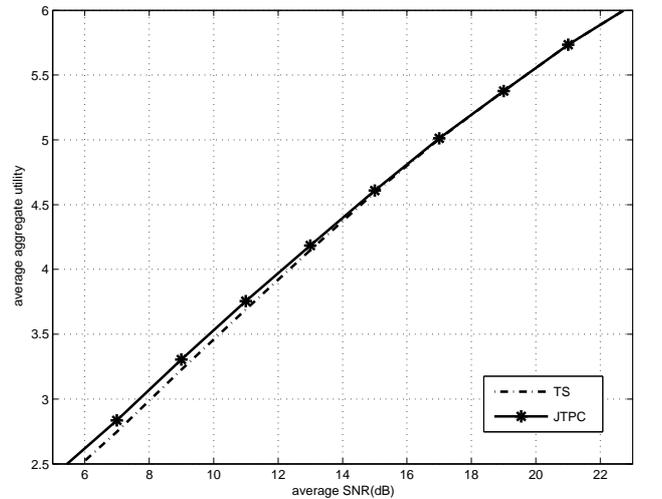} \caption{ Average aggregate utility versus average SNR
using TS and JTPA for two-user scenario} \label{fig_TS_JTPC}
\end{figure}

\begin{figure}
\centering
\includegraphics[width=3.3in]{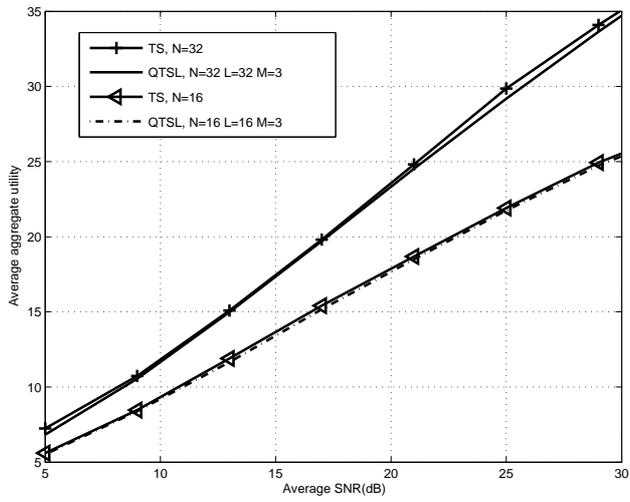}
\caption{Performance comparison of the optimal TS policy and QTSL
policy for $N=32$ and 16 users, with $L=N$ slots and $M=3$
feedback bits} \label{fig_QTSL1}
\end{figure}

\begin{figure}
\centering
\includegraphics[width=3.3in]{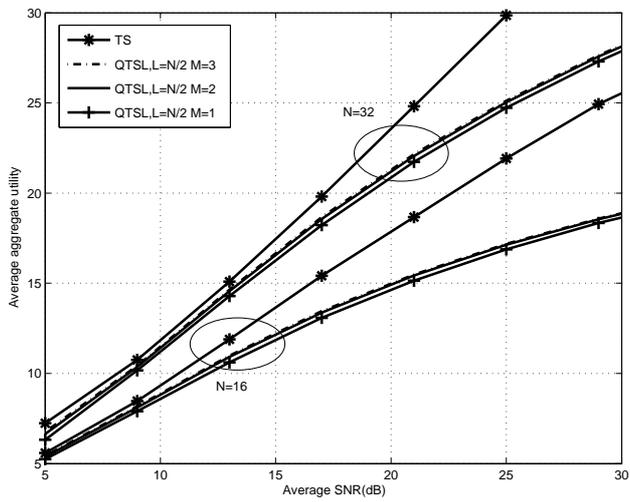}
\caption{Performance comparison of $N=32$ users with $L=16$ and
$N=16$ users with $L=8$ time slots, and with $M=1,2$ and 3
feedback bits } \label{fig_QTSL2}
\end{figure}

\end{document}